\documentclass{jpsj3}
\usepackage{graphicx,amsmath,amssymb,bm}

\allowdisplaybreaks 

\usepackage{color}
\definecolor{Green}{rgb}{0,0.7,0}

\newcommand{\ET}{ $\alpha$-(BEDT-TTF)$_2$I$_3$}

\newcommand{\bk}{ \bm{k}}
\newcommand{\bkD}{ \bm{k}_{\rm D}}
\newcommand{\eD}{ \epsilon_{\rm D}}
\newcommand{\ep}{ \epsilon }

\newcommand{\suz}[1]{\textcolor{black}{#1}}

\begin{document}

\title{
 Conductivity and Resistivity of Dirac Electrons
in Single-Component Molecular Conductor [Pd(dddt)$_2$]
}
\author{
Yoshikazu Suzumura\thanks{E-mail: suzumura@s.phys.nagoya-u.ac.jp}
$^{1}$
 HenBo Cui$^{2}$,   and 
 Reizo Kato$^{2}$
}
\inst{
$^1$
Department of Physics, Nagoya University,  Chikusa-ku, Nagoya 464-8602, Japan \\$^2$
RIKEN, 2-1 Hirosawa, Wako-shi, Saitama 351-0198, Japan \\
}

\recdate{February 6, 2018; accepted May 24, 2018}

\abst{
  Dirac electrons, which have been found in the single-component 
   molecular  conductor [Pd(dddt)$_2$] under  pressure, are examined  
 by calculating the  conductivity and resistivity 
 for  several pressures of $P$ GPa, 
   which give a nodal line semimetal or insulator.  
 The temperature ($T$) dependence of the conductivity is studied  
  using a tight-binding model with $P$-dependent transfer energies,  
 where the damping energy  by  the impurity scattering  $\Gamma$ is introduced.
 It is shown that  
 the conductivity  increases linearly under  pressure 
 at low $T$   due to the Dirac cone 
 but  stays almost  constant at high $T$. 
 Further,  at lower pressures,  the conductivity is suppressed  due to 
  an unconventional  gap,  which is examined by calculating the resistivity. 
  The resistivity exhibits a  pseudogap-like behavior
  even in the case  described by  the  Dirac cone.
  Such behavior originates  from a novel role  of the nodal line semimetal 
 followed by  a pseudogap  
 that is different from  a band gap. 
 The  present result  reasonably  explains  the resistivity 
 observed in the experiment.
  }


\maketitle

\section{Introduction} 
Massless Dirac fermions have recently  been extensively studied  due to their
  exotic energy band with a  Dirac cone. 
\cite{Novoselov2005_Nature438}
 Among them, Dirac electrons in molecular conductors
 have been  found in  bulk  systems\cite{Kajita_JPSJ2014,Kato_JACS}
 for the following two cases. 
  One is the organic conductor
     \ET 
      (BEDT-TTF=bis(ethylenedithio)tetrathiafulvalene), 
 in which a  zero gap is obtained in the two-dimensional Dirac electron
\cite{Katayama2006_JPSJ75} 
   using  the tight-binding model  
         with  the transfer energy estimated by 
         the extended H\"uckel method.\cite{Mori1984,Kondo2005} 
The other is the single-component molecular conductor 
  [Pd(dddt)$_2$] (dddt = 5,6-dihydro-1,4-dithiin-2,3-dithiolate)
 under a high pressure, in which   
 the  unconventional Dirac electron  was found  
  by  resistivity measurement\cite{Cui} and  first-principles calculation.
\cite{Tsumuraya_PSJ_2014}
 In fact, the latter material is a three-dimensional Dirac electron system 
 consisting of HOMO (highest occupied molecular orbital) and LUMO (lowest unoccupied molecular orbital) functions, and 
 exhibits  
      a  nodal line semimetal.\cite{Kato2017_JPSJ} 

Although there have been  many studies on  nodal line semimetals,
\cite{Murakami2007,Burkov2011,Yamakage2016_JPSJ} 
 the transport property has not been  studied  much except for that of 
 the molecular 
 conductors since the behavior relevant to 
 Dirac electrons can be obtained when  the  chemical potential is 
  located close to  the Dirac point. 
 One of the remarkable characteristics of  these molecular  Dirac electrons 
 is  the anisotropy of  the conductivity as shown  
  for both the two-dimensional case\cite{Katayama2006_cond}  
 and three-dimensional case.\cite{Suzumura2017_JPSJ} 
Although the conductivity at absolute zero temperature displays 
 a universal value  including the Planck constant, 
 the theory predicts an  increase in  the conductivity 
   when the temperature becomes larger than 
   $\Gamma$ associated with the energy by 
      the  impurity  scattering. 
 However,  the almost constant resistivity with varying temperature 
  is regarded  as  experimental evidence  of  Dirac electrons.
\cite{Tajima2007_EPL,Kato_JACS} 

 Since $\Gamma$ is much smaller than the energy of the Dirac cone, 
   it is not yet theoretically clear how to comprehend  
 the relevance between the Dirac electrons 
 and  the constant resistivity as a function of 
  temperature. 
Thus, in the present paper, we  examine 
 the temperature dependence of  both the  conductivity and resistivity 
 of the Dirac electrons of the single-component molecular conductor 
  [Pd(dddt)$_2$] under pressure 
  by taking into account the property of the actual band 
    through the transfer energy of the tight-binding model.\cite{Kato_JACS}
  Further, the characteristics of the pressure dependence 
     of the nodal line semimetal 
     are examined   using an interpolation formula for the transfer energy 
            between the ambient pressure and a pressure 
              corresponding to the observed Dirac electrons. 

In Sect. 2, the model is given 
  with the formulation for the conductivity. 
In Sect. 3, the nodal line semimetal under pressure is explained.   
 The temperature dependence of the anisotropic conductivity 
   is examined using the temperature dependence of the chemical potential, 
    which is determined self-consistently. 
  Further, the resistivity is calculated revealing  
  a pseudogap close to 
    the insulating state, and compared with the experimental result.
Section 4 is devoted to a summary and discussion.

\section{Model and Formulation}

 The single component molecular conductor [Pd(dddt)$_2$] \cite{Kato_JACS} 
  has a three-dimensional crystal structure with a unit cell 
 which   consists of four molecules (1, 2, 3, and 4)
   with HOMO and LUMO orbitals. 
 The molecules are located on two kinds of layers,  
 where layer 1 includes molecules 1 and 3 
     and  layer 2 includes  molecules 2 and 4.  
The transfer energies between nearest-neighbor molecules are given as follows. 
  The interlayer  energies in the $z$ direction  are given 
    by   $a$ ( molecules 1 and 2 and  molecules 3 and 4),
             and $c$ ( molecules 1 and 4 and  molecules 2 and 3). 
The intralayer  energies in the $x$-$y$ plane are given by 
   $p$ (molecules 1 and 3),  $q$ (molecules 2 and 4), 
    and   $b$  (perpendicular to the $x$-$z$ plane). 
Further, these energies are classified by 
 three  kinds of  transfer energies  given by   
   HOMO-HOMO (H), LUMO-LUMO (L),  and HOMO-LUMO (HL).

\begin{table}
\caption{ 
Transfer energies for $P$ = 8 and 0 GPa in the unit of eV. 
$b_1$, $p_1$, and $q_1$ 
($b_2$, $p_2$, and $q_2$) are the energies of layer 1 (layer 2), 
 where
  $p_{1H} = p_{2H} =p_{H}$, $q_{1H} = q_{2H} =q_{H}$, 
  $p_{1L} = p_{2L} =p_{L}$, and $q_{1L} = q_{2L} =q_{L}$. 
 }
\begin{center}
\begin{tabular} {ccccccccc}
\hline\noalign{\smallskip}
 $P=8$     & $a$   &  $b1$  & $b2$  & $c$ 
              & $p1$  &  $p2$  & $q1$  & $q2$  
\\
\noalign{\smallskip}\hline\noalign{\smallskip}
 $ H $     & $-0.0345$   &  $0.2040$  & $0.0762$  & $0.0118$ 
              & $0.0398$  &  $0.0398$  & $0.0247$  & $0.0247$  
\\
 $L $     & $0$   &  $0.0648$  & $-0.0413$  & $-0.0167$ 
              & $0.0205$  &  $0.0205$  & $0.0148$  & $0.0148$  
\\
 $HL$     & $0.0260$   &  $0.0219$  & $-0.0531$  & $0.0218$ 
              & $-0.0275$  &  $-0.0293$  & $-0.0186$  & $-0.0191$  
\\
\noalign{\smallskip}\hline
\hline\noalign{\smallskip}
 $P=0$     & $a$   &  $b1$  & $b2$  & $c$ 
              & $p1$  &  $p2$  & $q1$  & $q2$  
\\
\noalign{\smallskip}\hline\noalign{\smallskip}
 $ H $     & $-0.0136$   &  $0.112$  & $0.0647$  & $0$ 
              & $0.0102$  &  $0.0102$  & $0.0067$  & $0.0067$  
\\
 $L $     & $-0.0049$   &  $0.0198$  & $0$  & $-0.0031$ 
              & $0.0049$  &  $0.0049$  & $0.0037$  & $0.0037$  
\\
 $HL$     & $0.0104$   &  $0.0214$  & $-0.0219$  & $0.0040$ 
              & $-0.0067$  &  $-0.0074$  & $-0.0048$  & $-0.0051$  
\\
\noalign{\smallskip}\hline
\end{tabular}
\end{center}
\label{table_1}
\end{table}

Based on the crystal structure,\cite{Kato_JACS} 
  the tight-binding model Hamiltonian per spin is
  given by 
\begin{equation}
H = \sum_{i,j=1}^{N} \sum_{\alpha,\beta} t_{i,j;\alpha, \beta}(P) |i, \alpha> <j, \beta| \; ,
\label{eq:H_model}
\end{equation}
where  
 $t_{i,j;\alpha, \beta}$ are transfer energies between nearest-neighbor sites and $|i, \alpha>$ is a state vector. 
$i$ and $j$ are the  lattice sites  of the unit cell 
  with  $N$ being the total number of square lattices, 
     $\alpha$ and $\beta$ denote the eight molecular orbitals 
      given by the HOMO $(H1, H2, H3, H4)$ and 
         LUMO $(L1, L2, L3, L4)$.  
 The lattice constant is taken as unity. 
  For simplicity, the transfer energy  at  pressure $P$ (GPa) 
  is estimated by 
 linear interpolation between the two energies at  
  $P$ = 8  and 0 GPa.   
 With $r = P/P_0$ and $P_0= 8$ GPa, the energy is written as
\cite{Suzumura2017_JPSJ} 
\begin{equation}
t_{i,j; \alpha,\beta}(P)
  = r t_{i,j; \alpha,\beta}(0) + (1-r) t_{i,j; \alpha,\beta}(P_0) \; .
\label{eq:interpolation}
\end{equation}
 The insulating state is obtained at $P$=0 while 
the Dirac cone   is found at  $P_0$
by  first-principles calculation,\cite{Kato_JACS}  
 corresponding to the Dirac electron indicated by the experiment. 
 We take  eV as the unit of  energy. 
 The transfer energies $t_{i,j;\alpha, \beta}$ 
  at   $P$ = 8\cite{Kato_JACS} and   
 0 GPa\cite{Kato2017_JPSJ}), 
  which are obtained  by the extended  H\"uckel method,   
are listed in Table \ref{table_1}.
The gap 
 between the energy of the HOMO and that of the LUMO is taken 
 as $\Delta E = $ 0.696 eV to reproduce 
  the energy band in the first-principles calculation.

Using the Fourier transform 
$ |\alpha(\bm{k})>$ 
 $= N^{-1/2} \sum_{j} \exp[- i \bm{k}\bm{r}_j] \; |j,\alpha>$
       with the  wave vector  $\bk = (k_x, k_y, k_z)$,
Eq.~(\ref{eq:H_model}) is rewritten as
\begin{equation}
H = 
 \sum_{\bm{k}} |\Phi(\bm{k})> \hat{H}(\bm{k}) <\Phi(\bm{k})|\; , 
\label{eq:H} 
\end{equation}
 where
$<\Phi(\bm{k})| = (<H1|,<H2|,<H3|,<H4|, <L1|, <L2|, <L3|, <L4|)$ and 
is expressed as $<\Phi(\bm{k})| =  (<1|,<2|,\cdots, <8|.$  
 The Hermite  matrix Hamiltonian $\hat{H}(\bm{k})$
  is given in Ref.~\citen{Kato2017_JPSJ}, where  
 $\hat{H}(\bk)$ is expressed as $U(\bk)^{-1} H(\bk) U(\bk)$. The quantity 
 $H(\bk)$ [$U(\bk)$] is the real matrix Hamiltonian 
 (unitary matrix) shown in the Appendix. 
 The  nodal line 
   has been explained using  $H(\bk)$,\cite{Suzumura2016_JPSJ,Kato2017_JPSJ} 
 where the Dirac point is supported  by the existence of 
 an inversion center.\cite{Herring1937} 
 Note that  $\hat{H}(\bk)$ gives the same conductivity as $H(\bk)$ obtained from $U(\bk)$, which corresponds to a choice of the gauge. For simplicity,  
    the following calculation of the conductivity is performed 
  in terms of  $\hat{H}(\bk)$. 

The energy band  $E_j(\bk)$ 
 and the wave function $\Psi_j(\bk)$, $(j = 1, 2, \cdots, 8)$ 
 are calculated from 
\begin{equation}
\hat{H}(\bm{k}) \Psi_j(\bk) 
 = E_j(\bk) \Psi_j(\bk) \; , 
\label{eq:energy_band}
\end{equation}
 where $E_1 > E_2 > \cdots > E_8$ and 
\begin{equation}
\Psi_j(\bm{k}) = \sum_{\alpha}
 d_{j,\alpha}(\bk) |\alpha> \; ,
\label{eq:wave_function}
\end{equation}
 with $\alpha =$  
 $H1, H2, H3, H4, L1, L2, L3$, and $L4$. 
The Dirac point $\bk_D$ is obtained from 
 $ E_4(\bkD)= E_5(\bkD) \equiv  \ep (\bkD)$, 
 which gives a nodal line semimetal. 
The insulating state is obtained 
 for $E_G \not= 0$ 
  due to a half-filled band, 
 where  $E_{G} \equiv {\rm min} [E_4(\bk)-E_5(\bk)]$ 
 for all $\bk$.

Using $d_{\alpha \gamma}$  in Eq.~(\ref{eq:wave_function}), 
 the electric conductivity  per spin and per unit cell 
  is calculated as\cite{Katayama2006_cond,Suzumura2017_JPSJ}  
\begin{eqnarray}
\sigma_{\nu}(T) &=&  
   \int_{- \infty}^{\infty} d \omega
   \left( - \frac{\partial f(\omega) }{\partial \omega} \right) F_{\nu}(\omega) \; ,
   \label{eq:sigma} 
       \\
  F_{\nu}(\omega) &=&  
  \frac{e^2 }{\pi \hbar N} 
  \sum_{\bk} \sum_{\gamma, \gamma'} 
  v^\nu_{\gamma \gamma'}(\bk)^* 
  v^{\nu}_{\gamma' \gamma}(\bk) 
        \nonumber \\
 & &
 \times \frac{\Gamma}{(\omega - \xi_{\bk \gamma'})^2 + \Gamma^2} \times 
 \frac{\Gamma}{(\omega - \xi_{\bk \gamma})^2 +  \Gamma^2}
  \; ,  
   \nonumber \\ 
   \label{eq:Fz}
\\
  v^{\nu}_{\gamma \gamma'}(\bk)& = & \sum_{\alpha \beta}
 d_{\alpha \gamma}(\bk)^* 
   \frac{\partial \tilde{H}_{\alpha \beta}}{\partial k_{\nu}}
 d_{\beta \gamma'}(\bk) \; ,
  \label{eq:v}
\end{eqnarray}
 where $\nu = x, y,$ and $z$ and 
 $h = 2 \pi \hbar$. $h$ and $e$ denote
   the Planck constant and  electric charge,  respectively. 
 $\xi_{\bk \gamma} = E_{\gamma}(\bk) - \mu$ 
 and $\mu$ denotes the chemical potential. 
 $f(\omega)= 1/(\exp[\omega/T]+1)$ with $T$ being the temperature 
 in the unit of eV 
 and $k_{\rm B }=1$.
 The conductivity at absolute zero temperature was examined previously 
  by noting that $\sigma_{\nu}(0) = F_{\nu}(0)$.\cite{Suzumura2017_JPSJ} 
The energy $\Gamma$ due to the impurity scattering is introduced  
 to obtain a finite conductivity. 
The total number of  lattice sites  is given by $N=N_xN_yN_z$, where 
 $N_xN_y$  is the number of  intralayer sites and 
 $N_z$ is   the number of layers.   
Note that the calculation of Eq.~(\ref{eq:sigma}) 
 with the summation of $k_z$ at the end, i.e., 
 the two-dimensional conductivity  for a fixed $k_z$,  
  is useful to comprehend the nodal line semimetal 
     as shown previously.\cite{Suzumura2017_JPSJ} 

The chemical potential $\mu =  \mu(T)$ is determined self-consistently 
in the clean limit  from 
\begin{eqnarray}
  \frac{1}{N} \sum_{\bk} \sum_{\gamma}  f(E_{\gamma}(\bk) - \mu(T)) 
 & =& \int_{-\infty}^{\infty} {\rm d} \omega D(\omega) f(\omega) =  4 \; ,  
   \nonumber \\
  & & 
 \label{eq:eq11}
\end{eqnarray}
 which is  the half-filled condition 
 due to the HOMO and LUMO bands. 
 $D(\omega)$ denotes the density of states (DOS) per spin and per unit cell, 
 which is given by 
\begin{eqnarray}
D(\omega) &=& \frac{1}{N} \sum_{\bk} \sum_{\gamma}
 \delta (\omega - E_{\gamma}(\bk) + \mu) \; ,
  \label{eq:dos}
\end{eqnarray}
 where  $\int {\rm d} \omega D(\omega) = 8$.
 Note that Eq.~(\ref{eq:sigma}) can be understood 
 using the DOS when  the intraband contribution ($\gamma=\gamma'$)
  is dominant   and the $\bk$ dependence of 
 $v_{\gamma'\gamma}^{\nu}$ is small. 
 
\section{Conductivity and Resistivity of Nodal Line Semimetal in [Pd(dddt)$_2$]}

\begin{figure}
  \centering
\includegraphics[width=8cm]{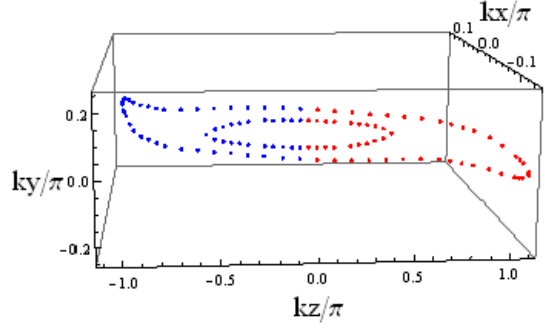}
    \caption{(Color online)
Nodal line of Dirac point at  $P$ = 7.7 GPa (inner loop) and 
 8.0 GPa (outer loop). 
 The former loop exists in the region of $|k_z/\pi| < 0.55$  while 
   the latter loop extends for arbitrary  $|k_z/\pi|$. 
 }
\label{fig:fig1}
\end{figure}

The electronic states of [Pd(dddt)$_2$] obtained 
 from the tight-binding model 
  show the following  pressure ($P$ GPa) dependence. 
 At ambient pressure ($P$ = 0), the insulating state is found  
 with  a  gap $E_g \simeq 0.41$ (eV)  at $\bk=(0,0,0)$,  which separates  
 the LUMO bands ($E_1,\cdots, E_4$) from the  HOMO bands ($E_5,\cdots, E_8$). 
With increasing $P$,  the gap decreases  and becomes  zero 
 at $P \simeq 7.58$, where  
  the Dirac point  given by  $E_4(\bk)= E_5(\bk)$ emerges at $\bk = 0$.
For $P > 7.58$,   the minimum of the LUMO band at $\bk = 0$ 
 becomes smaller than 
    the maximum of the HOMO band at $\bk=0$, 
 resulting in the following state.
As shown by
 the effective Hamiltonian 
 on the basis of the HOMO and LUMO orbitals,\cite{Kato2017_JPSJ} 
 a loop of the Dirac point    between  $E_4(\bm{k})$ (conduction band)  
    and  $E_5(\bm{k})$ (valence band)  emerges at the intersection 
 of the plane of   $E_4(\bk)= E_5(\bk)$ and that of  
 the vanishing of the H-L interaction (the coupling between HOMO  and LUMO orbitals).\cite{Kato_JACS,Kato2017_JPSJ} 
This loop gives a semimetallic state 
 since the chemical potential is located on a Dirac point of the loop
due to a half-filled band. 
 For $7.58 < P < 7.8$, the loop exists within the first Brillouin zone 
(as shown by the inner loop of Fig.~\ref{fig:fig1}).
In this case, 
  a gap for the fixed $k_z$ exists at  $\bk = (0,0,\pi)$ 
  but is absent at $\bk = (0,0,0)$.
 In fact, with decreasing $k_z$ from $\pi$, 
 the gap at $\bk = (0,0,k_z)$ given by 
   $E_{4}(0,0,k_z) - E_{5}(0,0,k_z)$  decreases and becomes 
 zero  at  $k_z/\pi \simeq$  0.55 in the case of  $P$=7.7.
 Such a gap is characteristic for the loop (i.e., nodal line) 
     within the first Brillouin zone.
 For  $P > 7.8$,   
 the loop exists in the extended zone (as shown by the outer loop of Fig.~\ref{fig:fig1}), and  the gap is absent for arbitrary $k_z$.
 Such a  gap for $7.57 < P < 7.8$
  could give rise to  novel behavior 
 in the  temperature dependence  of the conductivity as shown below. 
 
%

 Figure \ref{fig:fig1} 
  depicts  the nodal line of the loop of  the Dirac point 
 in the three-dimensional momentum space 
$\bk = (k_x, k_y,k_z)$ for $P$= 7.7 and  8,  where $P$=8  
 corresponds to   the pressure in the experiment 
  displaying  almost constant resistivity.\cite{Kato_JACS}  
In the previous work for $P$ = 8,\cite{Suzumura2017_JPSJ}
 it was shown that the Dirac cone is anisotropic,   the ratio of 
the velocity was estimated as 
 $v_x : v_y: v_z \simeq 1 : 5: 0.2$, 
 and that  
 the conductivity $\sigma_x$ 
 is always  relevant to  a  two-dimensional Dirac cone  
 rotating along the loop. 
Thus, we mainly study $\sigma_x$ as a typical conductivity 
 in the nodal line system. 
Note that the semimetallic state is obtained since the chemical potential 
 exists on the loop due to a half-filled band, for example, 
 $\mu$ (= 0.5561) 
 is located  on the loop with $|k_z/\pi| = 0.65$ for $P$ = 8. 
In fact, $\eD$ depends slightly on $k_z$ 
  to form an energy dispersion $\eD(k_z)$ with a width of $\simeq$ 0.003. 
Further, we note that the loop is not coplanar, 
 which is characteristic of the nodal line of the present system. 
 We take $e = \hbar$ = 1 in the following calculation. 

\begin{figure}
  \centering
\includegraphics[width=7cm]{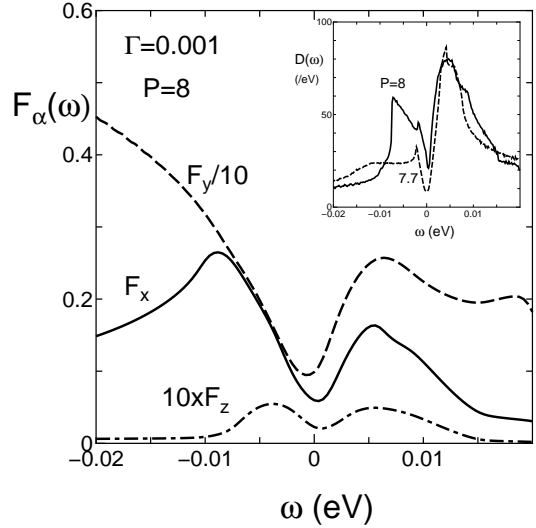}
     \caption{ 
Energy ($\omega$)  dependence of $F_{\alpha}$ ($\alpha$ = $x, y$, and $z$) 
 for $P$ = 8 (GPa) and $\Gamma$ = 0.001 (eV). 
The inset denotes the density of states (DOS) per spin
 for $P$ = 8  (solid line) and 7.7 (dashed line). 
$\omega = 0$ corresponds to the chemical potential at $T$ = 0, where 
 $\mu (0)$ = 0.5560 and 0.5558 for $P$ = 8 and 7.7, respectively.  
  }
\label{fig:fig2} 
\end{figure}

The conductivity is determined by $F_{\alpha}(\omega)$ ($\alpha$ = $x$, $y$, 
 and $z$)  of Eq.~(\ref{eq:Fz}), which is shown in Fig.~\ref{fig:fig2} 
 for $P$ = 8  and $\Gamma$ = 0.001. 
There are two peaks in $F_x$, where the difference in  magnitudes of 
 $F_\alpha$  originates  from  the factor  
$ v^{\nu}_{\gamma \gamma'}(\bk)$ in Eq.~(\ref{eq:v}).
 These peaks correspond to the top of the HOMO band for $\omega >0$ and the bottom of the LUMO band for $\omega < 0$.
 This can be  seen  from the DOS given by Eq.~(\ref{eq:dos}), 
 shown 
  in the inset  for  $P$ = 8 and 7.7,  
 where the 
  two peaks for  $\omega > 0$ and $\omega < 0$  come from the properties of 
 the band edges of the HOMO and LUMO, respectively. 
 Since the band  around the edge is strongly quasi-one-dimensional 
  along the $k_y$ direction, the peak is associated with the van Hove singularities of the quasi-one-dimensional band.   
The linear dependence around $\omega=0$ in the DOS is due to the Dirac cone,  
 and $D(0) \not= 0$ originates from  the semimetallic state 
 of the nodal line. 
The behavior of the Dirac electron is found for $|\omega| < 0.005$ eV, 
 implying that the $T$ dependence of the conductivity for the Dirac electron 
 is also expected for $T < 0.005$.  
The DOS outside of the peak becomes small  
 due to the energy of the conventional band. 
The present calculation is performed by choosing $\Gamma$ = 0.001, which 
 gives the following $T$ dependence. 
 The region $0 < T < 0.001$ corresponds to the semimetallic state. 
The Dirac electron gives  constant behavior of $\sigma_x$ 
 as a function of $T$  
 for $T < \Gamma =0.001$, while 
 $T$-linear dependence is expected in the 
 region of $0.001 < T < 0.005$.
For $0.005 < T$, the conductivity is determined by the effect  
 of the energy band outside of the Dirac cone.

\begin{figure}
  \centering
\includegraphics[width=7cm]{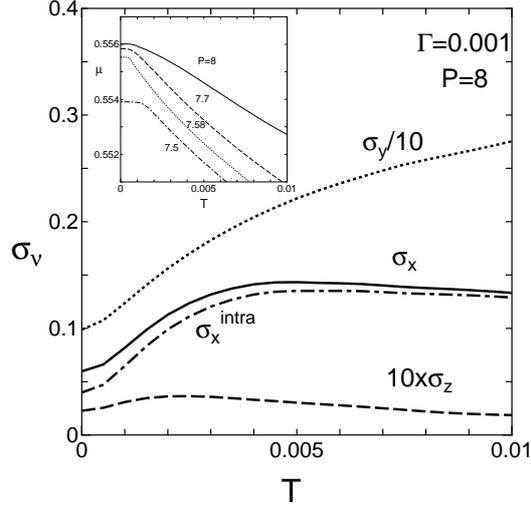} 
\caption{  
$T$ dependence of $\sigma_{\nu}$ ($\nu$ = $x$, $y$, and $z$) 
  for $P$ = 8 and $\Gamma$ = 0.001. 
The dot-dashed line denotes $\sigma_x^{\rm intra}$, which 
 corresponds to the intraband contribution.
The inset denotes $\mu(T)$ for $P$ = 8, 7.7, 5.58, and 7.5, which is 
 calculated from Eq.~(\ref{eq:eq11}).  
}
\label{fig:fig3}
\end{figure}
Figure \ref{fig:fig3} shows  $\sigma_{\nu}$ ($\nu$ = $x$, $y$, and $z$) 
 as a function of $T$ for $P$ = 8.
The inset denotes the $T$  dependence of $\mu (T)$,   
 which decreases monotonically,
 e.g., $\mu(0) - \mu(T) \sim 0.003$   at $T$ = 0.01  for $P$ = 8. 
Although such a chemical potential is replaced by an averaged one in the presence of a finite $\Gamma$,  the calculation of $\sigma_{\nu}$  
 with a chemical potential in the clean limit is reasonable for $T$ 
  much larger than $\Gamma$. 
Note that the $T$ dependence of  
$\sigma_\nu$ originates  from both 
 $f(\omega)$ in Eq.~(\ref{eq:sigma}) 
 and $\mu$ in $\xi_{\bk \gamma}$ of Eq.~(\ref{eq:Fz}). 
The anisotropy of $\sigma_{\mu}$ is large, where  
  both  $\sigma_x$ and $\sigma_z$ exhibit 
  a slight maximum      but $\sigma_y$ increases monotonically 
 with increasing $T$. 
 Typical behavior of the Dirac cone is seen  for 
  $\sigma_x$, which increases linearly  at low temperatures. 
For comparison,  
the intraband contribution  $\sigma_x^{\rm intra}$ 
 is shown by the dot-dashed line, which is obtained  for  $\gamma = \gamma'$. 
 The interband contribution is obtained by the  difference 
 between $\sigma_x$ (solid line) and  $\sigma_x^{\rm intra}$.  
 The increase in 
 $\sigma_x$ as a function of $T$ 
 is determined by the intraband contribution  
 while  $\sigma_x$ close to $T$=0  
  is determined by both intra- and interband contributions.
Since the maximum of $\sigma_x$  corresponds to 
 that of $F_{x}(\omega)$ in Fig.~\ref{fig:fig2}, 
 the  maximum suggests a 
 crossover from the Dirac electron to the conventional electron. 
We also calculated $\sigma_x$ by fixing $\mu$ at $T$ = 0 to see  the effect of the variation of $\mu$ on the conductivity. 
Such $\sigma_x$ is slightly smaller than that of the solid line 
 but is still located  above 
  the dot-dashed line in Fig.~\ref{fig:fig3}.  
 Thus, it turns out that  
  the increase in $\sigma_x$ at low temperatures 
    ($ 0.005 > T > \Gamma$) originates   
  from a property of the Dirac cone,    and  
  the almost $T$-independent $\sigma_x$ at higher $T$ $(> 0.005)$ is obtained 
 due to the suppression of the effect of the Dirac cone by the conventional band.

We note $\sigma_x(T)$ at $T$=0  
 in the presence of  both the  Dirac cone and the nodal line. 
Based on the previous calculation,\cite{Suzumura2014_JPSJ} 
 a simplified model of a Dirac cone with a chemical potential  $\tilde{\mu}$
   defined by an averaged $|\ep(k_z)|$    
   gives  $\sigma_x(0) \simeq 1/(2\pi^2) + (1/8\pi)\tilde{\mu}/\Gamma$, 
     where the first term corresponds to the 
     universal conductivity. 
Thus, the decrease in $\Gamma$ gives the increase in $\sigma_x(0)$, 
  which is also verified numerically. 

\begin{figure}
  \centering
\includegraphics[width=7cm]{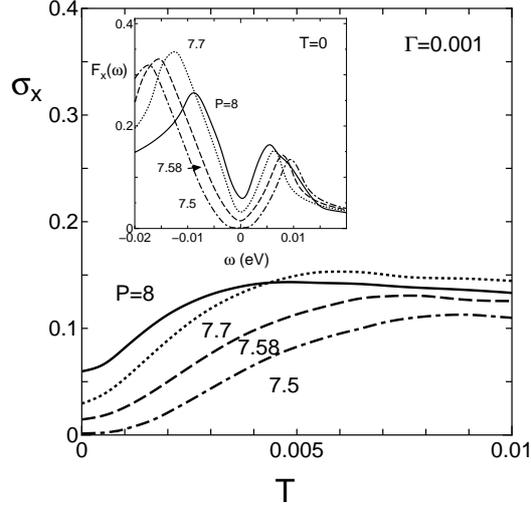}   
\caption{
$T$ dependence of $\sigma_x$ for $\Gamma$=0.001 
 with fixed $P$ = 7.5, 7.58, 7.7, and 8,  where the 
 $T$ dependence of $\mu (T)$ in Fig.~\ref{fig:fig3} 
 is taken into account.
The inset denotes the corresponding $F_x(z)$ at $T$ = 0. 
 }
\label{fig:fig4}
\end{figure}
In Fig.~\ref{fig:fig4}, the  $T$ dependence of $\sigma_x$  
 is examined  for  $P$ = 8, 7.7, 7.58, and 7.5.
The inset shows  the corresponding $F_x(\omega)$ at $T$=0.
Compared with $P$ = 8, the nodal line  for $P$=7.7 
 is reduced as shown  by  the inner loop of Fig.~\ref{fig:fig1}, since 
   a gap  exists  on  the $k_x$-$k_y$  plane  for a fixed $k_z$   
     with $0.55 < |k_z/\pi| \leq 1$. 
 Thus, $F_x(\omega)$ given by the dotted line in the inset decreases  
   for small $|\omega|$, leading to the suppression of  
     $\sigma_x$ (dotted line) at low temperatures. 
  This can also be  seen  from the DOS (inset of Fig.~\ref{fig:fig2}). 
 The linear increase in $\sigma_x$ in the case of  $P$ = 8 diminishes 
 and is replaced by  pseudogap behavior.
For $P$ = 7.58, corresponding to  the onset of the loop, 
the pseudogap behavior is further enhanced,  
 where the suppression  becomes larger but   
 $\sigma_x(0) \not= 0$ still remains due to  $\Gamma \not= 0$.
For $P$ = 7.5, $\sigma_x$ exhibits
  behavior expected in the insulating state. 
 The gap  at $\bk$ = 0 is estimated as $E_g \simeq 0.004$ from 
  the energy bands  $E_4(\bk)$ and $E_5(\bk)$. 
 The effect of $\Gamma$  still remains,  as seen from  
 the inset,  where  
 $F_x(\omega) \not= 0$  for $0.002 < |\omega| < E_g$ and  
 $F_x(\omega) \simeq  0$   for $|\omega| < 0.002$. 
Note that  $\sigma_x$  of the nodal line semimetal 
 for $P$ = 7.7 and 7.58   
 is characterized by such pseudogap behavior at lower temperatures  
 in addition to a slight maximum  
 at higher temperatures.

\begin{figure}
  \centering
\includegraphics[width=7cm]{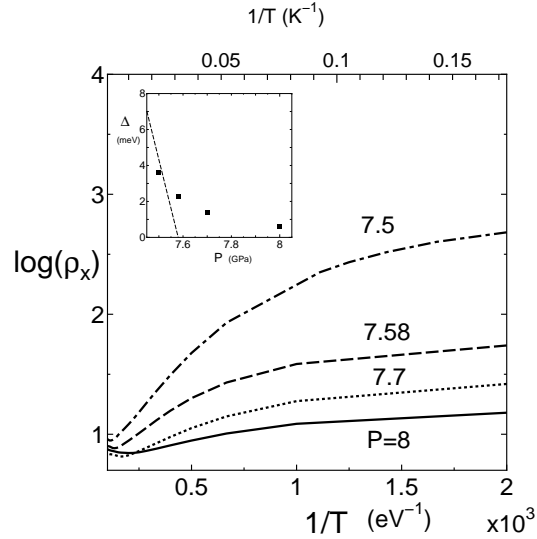}   
\caption{
  ${\rm Log} (\rho_x)-1/T$ plot (base 10 logarithm)
with $\rho_x=1/\sigma_x$ 
   for $P$ = 8, 7.7, 7.58, and 7.5.
The upper horizontal axis shows $1/T$ in the unit of $K^{-1}$ 
 for comparison with the experiment (Fig.~\ref{fig:fig6}). 
The inset shows the  $P$ dependence of 
 the pseudogap $\Delta$ in the unit of meV, which is estimated from the 
 tangent at $T \simeq 0.005$ in the main figure. 
 The dashed line denotes the band gap $E_G$ given 
 by $E_4(\bk) - E_5(\bk)$ at $\bk= 0$, which  decreases to zero at $P$ = 7.58. 
}
\label{fig:fig5}
\end{figure}

Now we examine the  behavior of $\sigma_x$ 
  for $T < 0.005$, which corresponds to  
    the region described by the Dirac cone. 
 Instead of $\sigma_x$, we calculate  the resistivity, which is obtained 
 by 
 $\rho_x=1/\sigma_x$ due to the  off-diagonal element being much smaller  than 
 the diagonal element.\cite{Suzumura2017_JPSJ} 
 The $T$ dependence of log($\rho_x$) (base 10 logarithm) 
   as a function of $1/T$ 
     is shown  in  Fig.~\ref{fig:fig5}  
         for $P$ = 8, 7.7, 7.58, and 7.5, the same values as  
 in Fig.~\ref{fig:fig4}. 
With decreasing $P$,  the increase in  ${\rm log}(\rho_x)$ 
  becomes steep due to the reduction of $D(0)$ 
    and the increase in the gap.  
For $P$=7.5, 
   behavior of the insulating state is seen, where 
  ${\rm log}(\rho_x) \propto 1/T$ suggests  an activation gap. 
 With increasing $P$,  this gap decreases while the following  
    characteristic appears as a nodal line semimetal. 
 The tangent of  ${\rm log}(\rho_x)$, 
   which is constant for small $1/T$,  begins to decrease  
  at $1/T \simeq 0.5 \times 10^3 \textendash 1 \times 10^3$ (eV)$^{-1}$ 
  and slowly varies in the region where  the energy 
 corresponds to the semimetal and $\Gamma$. 
 Thus, there are  two regions (I) and (II). 
 Gap behavior  occurs  in region (I) given by  
  $1/T < 0.5 \times 10^3$ (i.e., $0.002 < T < 0.005$) 
 while  semimetallic  behavior occurs 
 in  region (II) given by $ 10^3 < 1/T$ (i.e., $T < 0.001 = \Gamma$). 
We define  the gap $\Delta$ 
   by  $\partial ({\rm ln}( \rho) / \partial (1/T)$  
 in region (I). The gap $\Delta$ with some choices of $P$ 
 is shown in the inset, which is estimated from the main figure 
 in  Fig.~\ref{fig:fig5}  at 
 $1/T \simeq 0.005$ (eV)$^{-1}$ ($T \simeq 0.002$). 
 The dashed line in the inset denotes the band gap  $E_G$,   
 which is defined by ${\rm min} [E_4(\bk) - E_5(\bk)]$ 
 for arbitrary $\bk$. 
 With increasing $P$,
 $E_G $ ($ \simeq$ 0.041) at $P$ = 0  decreases linearly and becomes  
 zero at $P$ = 7.58.
  The gap $\Delta$ decreases as a function of $P$
     but deviates from the $P$-linear dependence 
    and remains finite even at $P$ = 8. 
 The gap $\Delta$ at $P$ = 7.5 corresponds  well to  $E_G$,  while 
   $\Delta \not= 0$  and $E_G =0$ at  $P$ = 7.58.
For $P$ = 7.7,
 the gap $\Delta$  in region (I) is 
  further reduced.
  The metallic contribution  in  region (II) 
    originates  from  the nodal line, 
     for $|k_z/\pi| < 0.55$, 
     which is enhanced by $\Gamma$, 
     as seen from $F_x(z)$ in  the inset of Fig.~\ref{fig:fig4}.  
 Thus,   pseudogap behavior is expected, i.e.,  
 the coexistence of the semimetallic component $|k_z/\pi| < 0.55$
 and the gap for $|k_z/\pi| > 0.55$.
For $P$ = 8, where the nodal line  exists 
  for arbitrary $k_z$, the crossover still exists but  
    the boundary  
  between region (I) and region (II) is invisible. 
Thus, such $P$ dependence of the gap $\Delta$ shows  
 a crossover from the insulating gap to the pseudogap 
 at $P \simeq 7.58$.

\begin{figure}
  \centering
\includegraphics[width=8cm]{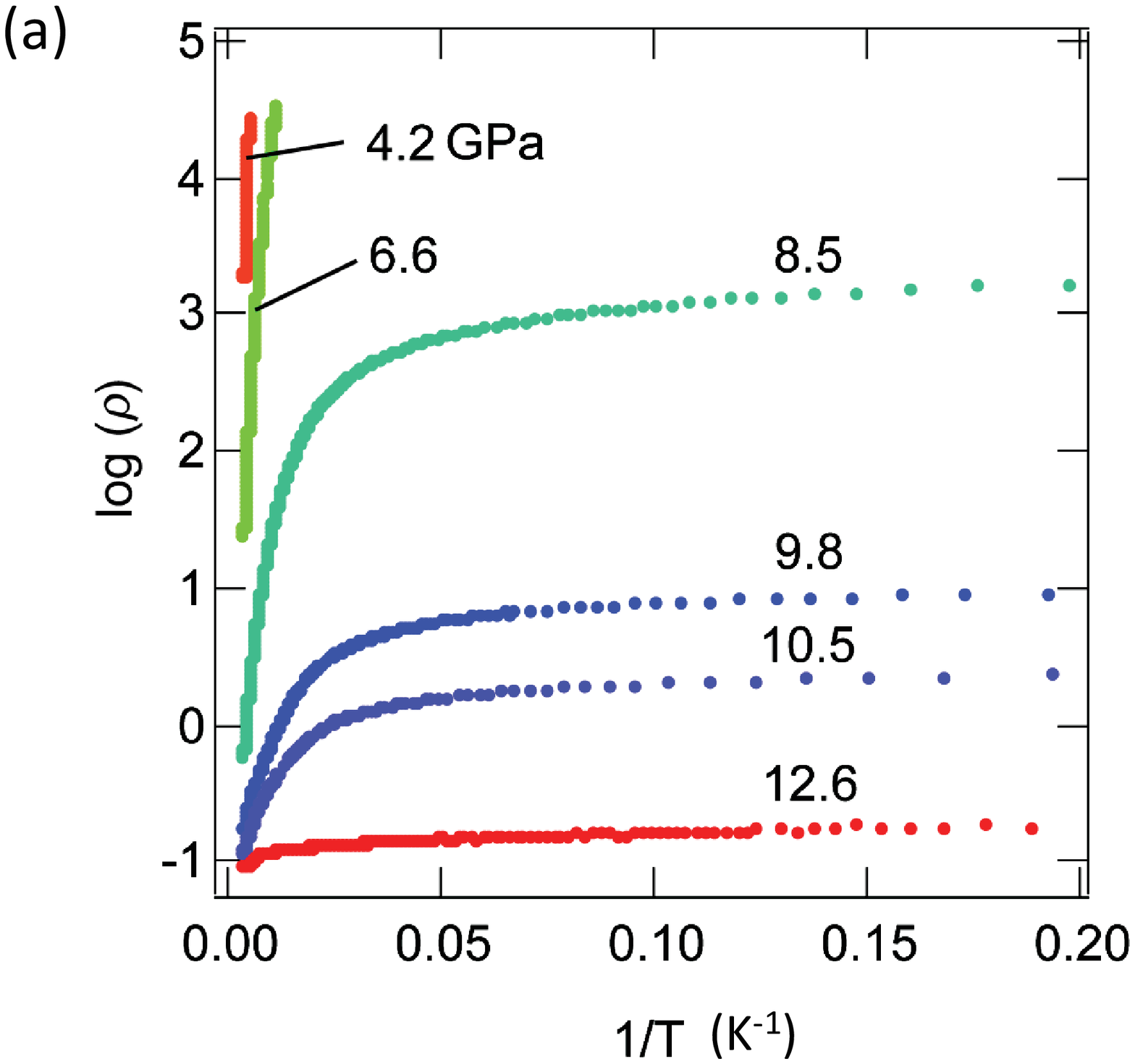}   
\includegraphics[width=8.5cm]{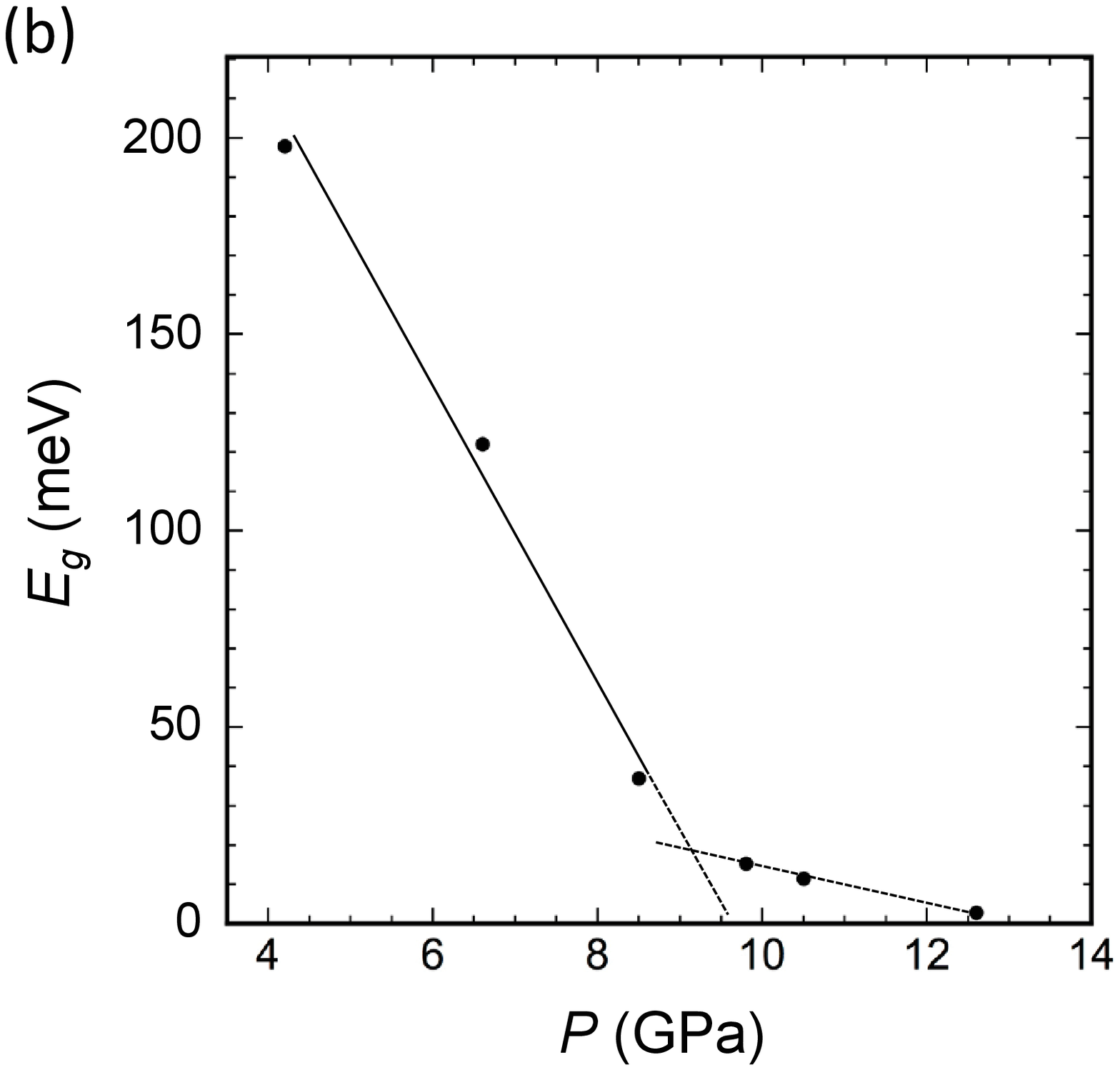}   
\caption{(Color online)
(a) log($\rho$) as a function of $1/T$, which is taken from 
 Ref.~\citen{Kato_JACS}, at 
 $P$ = 4.2, 6.6, 8.5, 9.8, 10.5, and 12.6 GPa,  
 where units are $\rho$ ($\Omega$ cm)  and $T$ (K).
(b) $P$ dependence of $E_g$ in the unit of meV, which is obtained from  the tangent of (a) 
 for  $T >$ 50 K,  where the lines are a guide to the eye.  
 }
\label{fig:fig6}
\end{figure}

Here we compare the present result of $\rho_x$ 
 with that observed in the experiment on [Pd(dddt)$_2$], 
 where the dc resistivity was measured along the longest side of the crystal (parallel to the $a+c$ direction). 
\cite{Kato_JACS}. 
Figure \ref{fig:fig6}(a), obtained from Ref.~\citen{Kato_JACS}, 
depicts the $T$ dependence of $\rho$,  
 where log($\rho$) is shown  as a function of  $1/T$ 
for several choices of $P$.  
These behaviors are also classified into two cases.
 For small $1/T$,
  the tangent of log($\rho$) as a function of $1/T$ 
  is large  [region (I)], whereas it is  
   small for large $1/T$ [region (II)]. 
 The temperature which separates region (I) from region (II) 
 is about $30 \textendash 50$ K for $P$ = 8.5, 9.8, 10.5, and 12.6 GPa.
 For  $P$ = 4.2 and 6,6 GPa,  $\rho$ is observed only 
  in region (I). 
 The gap $E_g$, which is obtained from the tangent in  region (I), 
 is shown in Fig.~\ref{fig:fig6}(b). 
 Two lines are drawn to guide the eye, where the line at lower pressures 
 indicates a band gap.  
 The gap at higher pressures    
  exhibits a $P$ dependence different from that  
    at lower pressures, suggesting a nontrivial origin. 
Figures \ref{fig:fig6}(a) and  \ref{fig:fig6}(b) are 
  compared with  Fig.~\ref{fig:fig5}. 
The $1/T$  dependence of log($\rho_x$) at  $P$=12.6 in the former 
   corresponds to 
     that of $P$ = 8  in the latter. 
 It is found that the crossover temperature between region (I) 
 and region (II) is comparable but the magnitude of $E_g$ in  the experiment 
 is much larger than $\Delta$ of the theory.
 The cases for $P$ = 9.8 and 10.5 GPa in Fig.~\ref{fig:fig6}(a) 
  suggest a state followed by 
  the pseudogap of the nodal line semimetal.

\section{Summary and Discussion}
We have examined the temperature dependence of the resistivity $\rho_x$ 
 using the conductivity $\sigma_x$ 
 for Dirac electrons in [Pd(dddt)$_2$], 
  which  exhibits a  nodal line semimetal. The main results are as follows. 

(i) For a pressure $P$ = 8 corresponding to the Dirac electron  observed 
 in the experiment,  
 a $T$-linear increase in $\sigma_x(T)$ is obtained   
  at low temperatures, whereas  $\sigma_x(T)$ 
    is almost constant   
      at high temperatures.  
 The former originates  from the Dirac cone and the latter is due to 
  the finite DOS of the conventional band outside of the Dirac cone. 
 The ratio $\sigma_x(0.001)/\sigma_x(0) \simeq 2 $ 
  obtained for $\Gamma$ = 0.001 is similar to that in 
 the experiment.\cite{Kato_JACS}   
 Note that  $\sigma_x$  depends on $\Gamma$
 since  the ratio $\sigma_x(0.001)/\sigma_x(0)$ 
 is given by 3.1, 2.0, 1.4, and 0.87 
 for $\Gamma$ = 0.0005, 0.001, 0.002, and 0.004, respectively, 
 suggesting a reasonable choice of $\Gamma$=0.001 
  to comprehend the results of the experiment. 
(ii) With decreasing $P$, $\sigma_x$ decreases due to the decrease in  
 the loop of the nodal line and the increase in the gap, as seen from  
  the calculation of   $\sigma_x$ for a fixed  $k_z$.
\cite{Suzumura2017_JPSJ}   
For $P$ = 7.7,  the nodal line exists for $k_z/\pi < 0.55$ and 
 is absent  for  $0.55 < k_z/\pi \leq 1$.
 This gives pseudogap behavior in $\sigma_x$  
 since the former gives a semimetal and the latter gives a gap.  
Such a magnitude of  the gap $\Delta$, which is estimated 
 from the tangent of the log($\rho$)-1/$T$ plot, 
 is different from the band gap $E_G$ and is characteristic 
 of a nodal line semimetal close to the insulating state. 
(iii)
The calculation of $\rho_x$  using the tight-binding model 
 is compared with that obtained experimentally  and 
   qualitatively good agreement is found. 
 The experimental results are understood 
  as  evidence of a nodal line semimetal, where 
  the loop of the Dirac point gives  pseudogap behavior 
 depending on the pressure.

We discuss  the agreement between the experiment and theory. Our tight-binding model with the transfer energies  well qualitatively describes the pressure and temperature dependence of the conductivity, but the determination of the absolute value is beyond the present scheme due to the H\"uckel approximation. 
For further comparison, it will be useful to observe the resistivity perpendicular to the present direction corresponding to the longest size of the crystal since  large anisotropy is predicted by the theory.

In addition to the energy band, the topological property of the wave function given by the Berry phase\cite{Berry1984} is important for understanding the Dirac  nodal line. The Berry curvature of two-dimensional Dirac electrons shows a peak around the Dirac point, which also occurs  for graphene and organic conductors.\cite{Fucks2010,Suzumura2011_JPSJ}  However, that of the nodal line is complicated due to the three-dimensional behavior, where the direction of the curvature rotates along the line. The Berry phase with a component parallel to the magnetic field could contribute to observe the Hall conductivity, which is expected to be different from that in two dimensions.\cite{Novoselov2005_Nature438,Tajima2013}

\acknowledgements
One of the authors thanks T. Tsumuraya and A. Yamakage for useful 
 discussions on the nodal line semimetal.  
This work was supported 
 by JSPS KAKENHI Grant Numbers JP15H02108 and JP16H06346.

\newpage

\appendix
 \section{Matrix elements of Hamiltonian}
We show explicitly the real matrix obtained  
 from  $\hat{H}(\bm{k})$ in Ref.~\citen{Kato2017_JPSJ}
 using   a  unitary transformation 
 given by 
$(U \hat{H} U^{-1})_{ij} = h_{i,j}$,  
 where 
 $U$ has only the diagonal matrix elements given by 
$(U)_{11} = -i$, 
$(U)_{22} = -ie^{-i(x+y+z)/2}$, 
$(U)_{33} = -ie^{-i(x+y)/2}$, 
$(U)_{44} = -ie^{iz/2}$, 
$(U)_{55} = 1$, 
$(U)_{66} = e^{-i(x+y+z)/2}$, 
$(U)_{77} = e^{-i(x+y)/2}$, and 
$(U)_{88} = e^{iz/2}$ with   
 $x=k_x, y=k_y$, and $z=k_z$.
  Since the symmetry of the HOMO (LUMO) 
 is odd (even)  with respect to the 
 Pd atom, the matrix elements of H-L ($h_{i,j}$ with 
 $i=1,\cdots, 4$, and $j = 5,\cdots 8$) 
 are  odd  functions 
 with respect to $\bm{k}$, i.e., 
  antisymmetric  at the time-reversal-invariant momentum.

 The real matrix elements $h_{ij}$ ($i,j =1, \cdots 8$)
 are shown in  Table \ref{table_2}, 
 where   
 the matrix elements of $h_{i,j}$  are  divided into the 4 x 4 matrices, 
  $h_{H,H}$, $h_{H,L}$, $h_{L,L}$ corresponding to the  H-H, H-L and L-L 
 components, respectively.   
  The functions in Table \ref{table_2} are defined by   
 $c(a,b)= 2 a \cos (b/2)$,  
 $s(a,b)= 2 a \sin (b/2)$, 
 $c_1(a)= c(a,x+y)+c(a,x-y)$,  
 $s_1(a)= s(a_{1HL},x+y)+s(a_{2HL},x-y)$, and  
 $s_2(a)= s(a_{1HL},x-y)+s(a_{2HL},x+y)$ 
 with 
 $x=k_x$, $y=k_y$, and $z=k_z$,  
 where  * denotes $h_{j,i} = h_{i,j}$.

\begin{table}
\caption{ Matrix elements of real Hamiltonian $ h_{i,j}$. 
 }
\begin{center}
\begin{tabular} {ccccc}
\hline\noalign{\smallskip}
$h_{H,H}$        & $H1$ & $H2$ & $H3$ & $H4$ 
\\
\noalign{\smallskip}\hline\noalign{\smallskip}
$ H1$   & $\suz{c(b_{1H},2y)}$  & $\suz{c(a_{H},x+y+z)}$  & $\suz{c_1(p_{H})}$  & $\suz{c(c_{H},z)}$  
  \\
$ H2$ 
 & $\suz{*}$  & $\suz{c(b_{2H},2y)}$  & $\suz{c(c{H},z)}$  & $\suz{c_1(q_{H})}$   
\\
$ H3$ 
 & $*$  & $*$  &  $ c(b_{1H},2y)$  & $c(a_{H},x-y+z) $   
\\
$ H4$ 
 & $*$  & $*$  &  $*$  & $c(b_{2H},2y) $   
\\
\noalign{\smallskip}\hline
\hline\noalign{\smallskip}
$h_{H,L}$        & $L1$ & $L2$ & $L3$ & $L4$ 
\\
\noalign{\smallskip}\hline\noalign{\smallskip}
$ H1$ 
  & $s(b_{1HL},2y)$  & $0$  & $s_1(p)$  & $0$   
  \\
$ H2$ 
  & $-s(a_{HL},x+y+z)$  & $s(b_{2HL},2y)$  & $-s(c_{HL},2y+z)$  & $s_2(q)$   
\\
$ H3$ 
  & $-s_2(p)$  & $0$  & $s(b_{1HL},2y)$  & $0$   
  \\
$ H4$ 
  & $-s(c_{HL},2y-z)$  & $-s_1(q)$  & $s(a_{HL},x-y+z)$  & $s(b_{2HL},2y)$   
\\
\noalign{\smallskip}\hline
\hline\noalign{\smallskip}
$h_{L,L}$        & $L1$ & $L2$ & $L3$ & $L4$ 
\\
\noalign{\smallskip}\hline\noalign{\smallskip}
$ L1$ 
  & $\Delta E + c(b_{1L},2y)$  & $ c(a_{L},x+y+z)$  & $ c_1(p_{L})$  & $c(c_{L},2y-z)$   
  \\
$ L2$ 
  & $*$  & $ \Delta E + c(b_{2L},2y)$  & $c(c_{L},2y+z)$  & $c_1(q_{L})$   
  \\
$ L3$ 
  & $*$  & $*$  & $\Delta E + c(b_{1L},2y)$  & $c(a_{L},x-y+z)$   
  \\
$ L4$ 
  & $*$  & $*$  & $*$  & $\Delta E + c(b_{2L},2y)$   
  \\
\noalign{\smallskip}\hline
\end{tabular}
\end{center}
\label{table_2}
\end{table}


\noindent Note added in proof 

\noindent We noticed that another group [Z. Liu, H. Wang, Z. F. Wang, J. Yang, and F. Liu, 
 Phys. Rev. B \textbf{97},  155138 (2018)]  performed the first-principles band calculation using our structural data\cite{Kato_JACS} 
and reconfirmed the nodal line semimetal character of [Pd(dddt)$_2$].


\end{document}